# Random Sequences from Primitive Pythagorean Triples

Monisha Prabhu and Subhash Kak

**Abstract:** This paper shows that the six classes of PPTs can be put into two groups. Autocorrelation and cross-correlation functions of the six classes derived from the gaps between each class type have been computed. It is shown that Classes A and D (in which the largest term is divisible by 5) are different from the other four classes in their randomness properties if they are ordered by the largest term. In the other two orderings each of the six random Baudhāyana sequences has excellent randomness properties.

**Introduction**

This article extends the results of a previous article on primitive Pythagorean triples (PPTs) [1], which presented their historical background and some advanced properties. This research is part of a program to use mathematical functions to generate random numbers [2]-[9].

There has been much recent research on families of Pythagorean triples [10]-[12]. A Pythagorean triple (*a, b, c*) is a set of integers that are the sides of a right triangle and thus $a^2 + b^2 = c^2$. Given a Pythagorean triple (*a, b, c*), (*da, db, dc*) is also a triple. A primitive Pythagorean triple (PPT) consists of numbers that are relatively prime.

To generate PPTs, one may start with different odd integers *s* and *t* that have no common factors and compute:

$$a = st; b = \frac{s^2 - t^2}{2}; c = \frac{s^2 + t^2}{2}$$

There exist an infinity of PPTs. The coordinate *(a/c,b/c)* may be seen as a point on the unit circle, implying that a countably infinity of these points are rational. A sequence that generates a subset of PPTs is *(2n+1, 2n²+2n, 2n²+2n+1)* for *n = 1,2,3…*

**Indexing Using s and t Numbers**

For a convenient indexing one may use relatively prime *s* and *t* numbers in an array where *s > t*. This may be seen in the diagram shown below:

|   |    | *s* |     |     |     |     |     |   |   |   |
|---|----|-----|-----|-----|-----|-----|-----|---|---|---|
|   |    | 3   | 5   | 7   | 9   | 11  | 13  | . | . | . |
|   | 1  | (1,3)| (1,5)| (1,7)| (1,9)| (1,11)| (1,13)| . | . | . |
|   | 3  |     | (3,5)| (3,7)| (3,9)| (3,11)| (3,13)| . | . | . |
|   | 5  |     |     | (5,7)| (5,9)| (5,11)| (5,13)| . | . | . |
| *t* | 7  |     |     |     | (7,9)| (7,11)| (7,13)| . | . | . |
|   | 9  |     |     |     |     | (9,11)| (9,13)| . | . | . |
|   | 11 |     |     |     |     |     | (11,13)| . | . | . |
|   | .  |     |     |     |     |     |     | . | . | . |
|   | .  |     |     |     |     |     |     |   |   |   |

**Table 1.** Array of PPTs ordered by *s* and *t* numbers

If the indexing were done according to columns of Table 1, we have the following PPTs in the array across the first seven generations (where the duplicate values have been removed):

```
(3,4,5)
(5,12,13)     (15,8,17)
(7,24,25)     (21,20,29)     (35,12,37)
(9,40,41)     (45,28,53)     (63,16,65)
(11,60,61)    (33,56,65)     (55,48,73)    (77,36,85)     (99,20,101)
(13,84,85)    (39,80,89)     (65,72,97)    (91,60,109)    (117,44,125)   (143,24,145)
(15,112,113)  (105,88,137)   (165,52,173)  (195,28,197)
  .    .    .    .    .    .    .
  .    .    .    .    .    .    .
```
**Table 2.** Example generations of PPTs ordered by *s* and *t* numbers

The fourth row has only three entries as (2,9) of Table 1, which corresponds to the triple (27,36,45) can be reduced to the PPT (3,4,5).

Indexing of the PPTs in Table 2 may be done according to increasing *a, b,* or *c*.

**Six Classes of PPTs**

**Theorem.** *Primitive Pythagorean triples come in 6 classes based on the divisibility of a, b, c by 3, 4, and 5.*

*Proof.* See [1]. The six classes are defined as follows:

1. Class A, in which *a* is divisible by 3 and *c* is divisible by 5. Examples: (3,4,5), (33,56,65)
2. Class B, in which *a* is divisible by 5, and *b* is divisible by 3. Examples: (5,12,13), (35,12,37).
3. Class C, in which *a* is divisible by 3 and 5. Examples: (15,8,17), (45,28,53)
4. Class D, in which *b* is divisible by 3 and *c* is divisible by 5. Examples: (7,24,25), (13,84,85)
5. Class E, in which *a* is divisible by 3 and *b* is divisible by 5. Examples: (21,20,29), (9,40,41)
6. Class F, in which b is divisible by 3 and 5. Examples: (11,60,61), (91,60,109)

The six classes may also be shown to be defined as the end nodes of the binary branching tree of Figure 1.

An experiment was done where 4448 PPTs were generated and indexed by increasing a, *b,* and *c*, respectively.

*Indexed by increasing a:*

>ABDEFDCCDFEEDBAFFAABBDEEFDCCDFEEDBBAAFFAABBDEEFDCCDDFEDBBAAFFFAAB
BDEEFDCCCCDFEEDBBAAFFFAABDEEFDDCCDFEEDDBBAAFFAABBDEEFFDCCCCDFEEDB
BAAFFAABB…

*Indexed by increasing b:*

>ACBBAEEDDCCCDDEEAABBCCAAFFFFACCBBAAEEDDDDCCCCDDEEAABBCCAAFFFFAAC
BBBBAAEEEEDDCCCCDDDDEEAABBBBCCAAFFFFAACCBBAAEEDDDDCCCCDDEEEEAABB
BBCCAAFF…

*Indexed by increasing c:*

>ABCDEBECFAABDDEBEFCACDDEBFCFAABCDDEEFCFCDDEBEFCAABCDDBFCAABCCEBEF
FAABDDEBEFFAABBDDEECCFAACDDEBCFBCDEEECFBCDDEEBBEFAABDDBEFCAABBCEE
FFAABCDD…

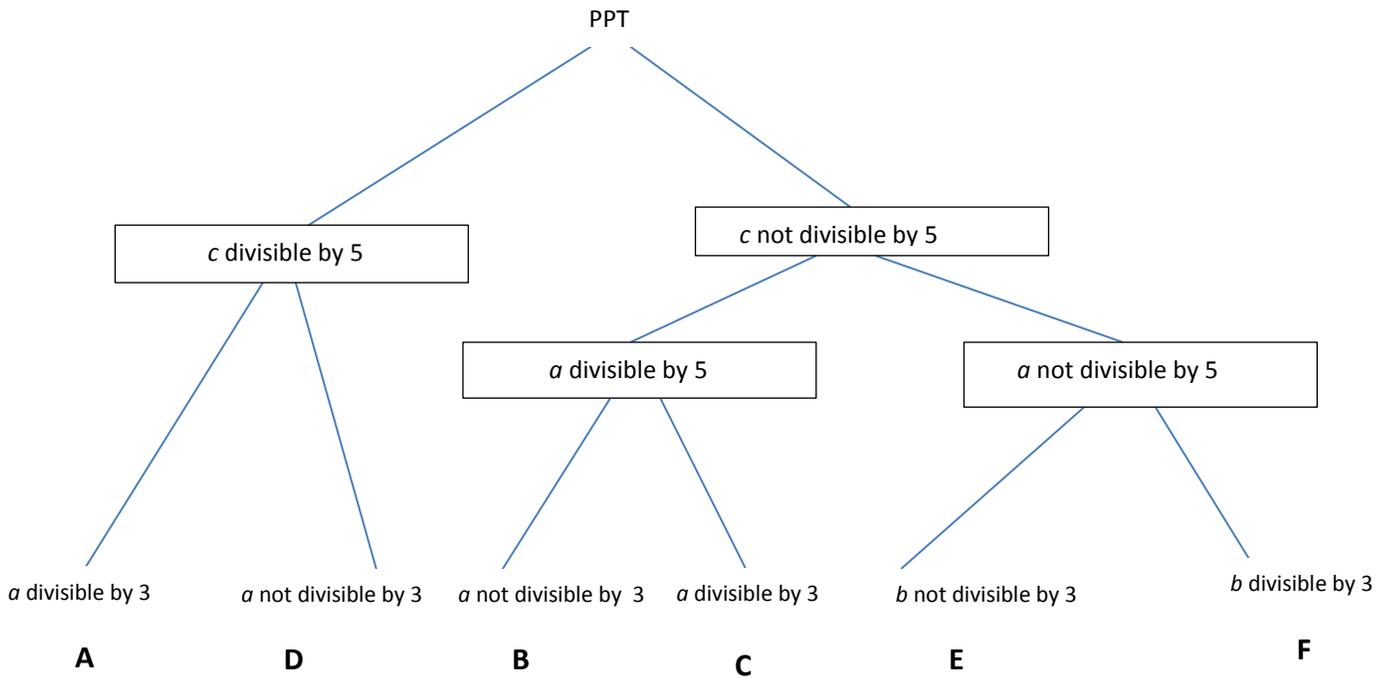

**Figure 1**. A graphical representation of the 6 classes of PPTs

**Randomness Properties of Sequences from the Six Classes**

We obtain separate sequences related to the occurrence of As, Bs, Cs, Ds, Es, and Fs by considering the distance between occurrences of the letters. Thus in the listing by increasing *c*, A occurs, after its first value, at the $10^{th}$, $11^{th}$, $20^{th}$, .. positions, which corresponds to the numbers 9, 1, 9, …. These sequences will be called *Baudhāyana sequences* after the author who used Pythagorean triples several centuries before Pythagoras [13],[14].

*Baudhāyana sequences ordered by a:*

Sequence of As: 14 3 1 17 1 3 1 17 1 4 1 19 1 4 1 18 1 3 1 20 1 3 1 18 2 10 11 1 1 1 9 3 10 2 9 1 1 1 10 2 11 1 12 2 12 2 1 1 10 2 11 2 13 1 13 1 9 1 11 1 1 1 10 1 1 21 11 1 3 21 12 8 1 3 1 9 11 1 2 10 1 12 1 1 9 18 2 1 9 …

Sequence of Bs: 12 6 1 13 1 7 1 13 1 8 1 15 1 8 15 1 7 1 16 1 7 1 15 4 8 2 9 5 7 5 8 4 7 5 6 1 1 4 9 3 10 4 1 1 8 6 8 4 9 4 11 3 9 1 1 3 11 8 4 1 1 5 5 8 1 1 2 7 3 8 6 8 1 1 1 8 2 8 4 6 7 7 3 1 1 6 5 8 4 8 1 4 6 1 9 3 1 4 1 26 ...

Sequence of Cs: 1 19 1 21 1 22 1 1 1 22 1 23 1 1 1 22 1 1 1 13 10 1 1 10 1 1 10 1 1 11 1 1 1 10 14 12 1 1 11 1 1 13 1 1 10 1 1 12 1 1 12 1 1 12 1 9 1 13 10 1 10 1 11 1 1 11 1 12 1 7 1 1 8 1 14 1 10 1 1 11 1 13 9 1 7 12 9 9 …

Sequence of Ds: 3 3 4 9 4 3 4 11 4 3 1 3 12 4 5 4 11 4 1 3 4 1 11 5 5 4 11 1 4 5 3 6 10 2 16 14 21 12 2 2 5 7 15 9 4 7 5 2 8 14 18 5 14 10 1 3 2 9 9 4 9 5 5 6 4 16 10 3 1 7 13 14 2 7 3 1 4 2 4 17 3 4 6 7 2 7 14 2 2 4 3 17 …

Sequence of Es: 7 1 11 1 7 1 13 1 8 14 1 9 1 13 1 8 1 14 1 10 1 14 1 9 8 3 1 1 5 6 7 5 7 5 7 5 7 4 8 1 1 4 7 1 1 5 8 5 9 6 7 5 7 7 1 1 5 7 7 1 4 7 4 1 23 11 1 5 1 18 7 1 1 12 1 9 10 23 1 5 6 24 1 1 7 12 1 13 6 9 1 3 1 5 1 5 6 …

Sequence of Fs: 5 6 1 8 5 8 1 8 6 7 1 1 8 7 8 1 1 7 6 9 1 8 1 7 8 1 9 7 5 5 14 16 1 7 5 4 12 8 26 9 5 14 13 4 5 9 5 10 7 6 11 2 2 15 2 8 12 1 3 4 18 7 1 5 6 1 12 8 4 8 5 3 1 1 26 1 5 4 3 6 1 9 2 8 4 4 1 9 6 5 13 1 1 9 3 1 13 2 7 …

*Baudhāyana sequences ordered by b:*

Sequence of As: 4 12 1 5 1 5 5 1 15 1 5 1 5 1 6 1 17 1 7 1 5 1 5 1 17 1 7 1 5 1 7 21 1 5 1 5 1 9 1 17 1 7 1 5 1 1 1 7 1 16 5 8 3 15 5 1 1 4 5 12 7 4 5 15 5 4 5 1 1 13 1 1 3 8 5 15 5 4 7 17 3 8 16 4 1 11 12 1 4 3 19 10 17 10 1 12 1 1 14 15 …

Sequence of Bs: 1 15 1 12 1 19 1 12 1 1 1 21 1 1 1 13 1 21 1 1 1 13 1 1 1 24 1 15 1 1 1 21 1 1 1 15 1 1 1 19 1 1 12 17 1 1 10 1 1 14 1 1 10 1 1 17 1 1 8 1 1 19 12 1 1 17 1 1 10 1 1 19 12 1 1 13 1 12 1 15 1 1 5 1 1 17 1 20 1 1 10 1 1...

Sequence of Cs: 8 1 1 9 1 8 1 11 1 1 1 9 1 9 13 1 1 1 13 1 9 1 11 1 1 1 13 1 9 1 14 1 1 1 13 1 9 1 1 1 11 1 1 1 15 1 11 1 13 1 1 1 10 10 9 1 1 8 9 11 1 1 6 11 1 1 1 9 6 13 1 9 10 11 1 1 6 1 1 11 1 1 1 9 10 7 1 1 9 10 7 1 1 27 9 1 1 4 …

Sequence of Ds: 1 4 1 24 1 1 1 5 1 28 1 5 1 1 1 27 1 1 1 5 1 32 1 1 1 5 1 1 1 31 1 5 1 1 1 35 1 5 1 1 1 22 1 1 3 1 1 22 1 1 2 1 1 24 1 1 5 24 1 1 3 1 1 24 1 1 3 1 1 24 1 1 5 1 1 21 1 4 1 1 22 4 1 19 1 1 1 1 2 1 1 16 1 1 4 1 22 1 2 1 1 22 …

Sequence of Es: 1 8 1 20 1 11 1 22 1 1 1 11 1 23 1 11 1 1 1 24 1 1 1 13 1 1 1 25 1 11 1 1 1 27 1 1 1 11 18 1 1 9 1 1 18 8 1 1 18 1 1 9 1 1 18 1 1 9 20 1 1 9 1 1 18 1 1 1 11 1 1 17 9 1 18 1 7 1 1 13 1 1 1 11 1 12 1 9 18 1 1 7 1 18 1 7 1 1 17 1 1 …

Sequence of Fs: 1 1 1 30 1 1 1 36 1 1 1 35 1 1 1 38 1 1 1 39 1 1 1 34 1 1 1 1 1 27 1 1 26 1 1 27 1 1 27 1 1 1 1 1 1 27 1 1 29 1 1 1 1 1 1 23 1 1 1 1 1 1 22 1 26 1 1 20 1 1 1 1 1 23 1 1 24 1 1 1 1 1 24 1 21 1 1 1 1 1 21 1 1 1 1 1 20 1 1 1 …

*Baudhāyana sequences ordered by c:*

Sequence of As: 9 1 9 9 1 18 1 8 1 9 1 9 1 10 1 26 1 8 1 8 1 9 1 9 1 8 1 18 1 11 1 19 1 7 1 8 1 19 1 8 1 7 1 9 1 16 1 21 1 9 1 5 1 11 1 6 1 11 1 17 1 1 1 6 1 1 1 8 1 30 1 17 1 9 1 18 1 7 1 8 31 1 8 1 6 1 10 1 1 1 9 1 17 1 7 …

Sequence of Bs: 4 6 4 9 6 13 6 4 5 4 6 4 6 1 14 3 8 6 1 5 3 6 1 8 4 9 10 6 8 7 4 12 3 10 18 1 5 8 1 5 1 7 10 5 3 1 6 5 12 11 11 4 6 6 6 6 1 1 10 1 6 9 1 2 11 6 9 1 4 5 4 10 7 11 7 4 8 5 5 1 8 14 1 13 5 4 1 8 9 7 6 11 13 4 7 2...

Sequence of Cs: 5 11 2 6 5 6 2 7 4 5 4 1 24 1 4 5 3 5 3 17 5 8 16 3 5 5 1 4 1 2 11 1 6 5 8 6 1 8 10 10 3 1 6 8 4 9 5 16 1 5 7 4 1 6 4 16 1 6 7 4 8 5 6 4 4 5 10 24 1 9 11 1 7 1 4 5 1 3 8 6 1 12 1 9 15 1 2 7 1 17 16 1 5 4 4 4 6 …

Sequence of Ds: 9 1 8 1 10 1 7 1 10 1 17 1 10 1 9 1 7 8 1 10 1 18 1 7 1 10 1 17 1 1 1 9 1 18 1 9 1 7 1 9 1 10 1 7 1 27 1 8 1 8 1 1 1 9 1 9 1 16 1 16 1 12 1 18 1 18 1 8 1 6 1 1 1 11 1 6 1 11 1 16 1 19 1 8 1 1 1 5 1 9 1 20 1 …

Sequence of Es: 2 8 2 7 11 1 7 2 17 2 8 2 9 1 9 7 1 1 7 1 3 8 8 1 10 7 2 18 1 10 10 9 2 1 7 2 8 8 10 1 1 18 7 1 8 3 8 1 2 1 16 1 9 1 1 15 1 9 2 15 1 12 9 11 7 1 10 2 1 6 2 7 1 11 1 6 2 11 6 10 8 1 2 9 11 1 5 2 10 1 7 1 1 9 9 …

Sequence of Fs: 9 8 2 9 2 7 9 10 1 9 1 11 9 8 10 8 9 1 9 1 8 1 8 2 8 11 9 1 2 9 1 1 9 6 2 8 10 10 8 2 7 1 10 11 1 5 12 1 17 1 1 6 12 7 10 2 8 1 2 6 19 1 1 11 1 16 2 1 8 8 18 1 1 1 10 7 1 8 1 11 1 6 13 7 1 1 18 10 2 5 1 2 1 …

We consider the autocorrelation function, *C(k),* of these sequences when written as *a(i)*:

$$C(k) = \frac{1}{n} \sum_{i=1}^{n} a(i)a(i+k)$$

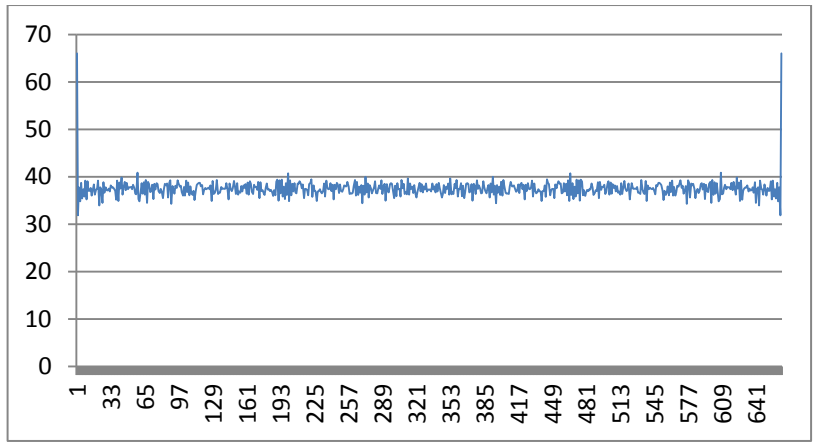

**Figure 2.** The typical autocorrelation function for Baudhāyana sequences ordered by *a* and *b*

The autocorrelation function for each of the six classes of Baudhāyana sequences ordered by *a* and *b* is qualitatively the same (Figure 2). Since there are 6 classes, the average distance between consecutive points will be 6. This, in turn, makes the function for non-zero values of the argument to be approximately 36 as we find in the plots.

The autocorrelation functions of the six random Baudhāyana sequences for ordering by *c* is shown in Figure 3. Notice that the value of the autocorrelation function for zero lag is not the same for all sequences. The functions for classes B, C, E, and F are similar to the results in Figure 2. However when the Baudhāyana sequences are arranged by order of *c*, As and Ds bunch together as there are more than one solution for values of *c* that are divisible by 5. These are represented by classes A and D. This is the reason why the plots for these two classes are different from the others as shown in Figure 3. This means that the correlation of Baudhāyana sequences ordered by c as shown in Figure 3a and Figure 3d is an artifact of the ordering process and the six sequences have excellent randomness properties if we order them by *a* or *b* or if we consider the classes B,C, E, and F when they are ordered by *c*.

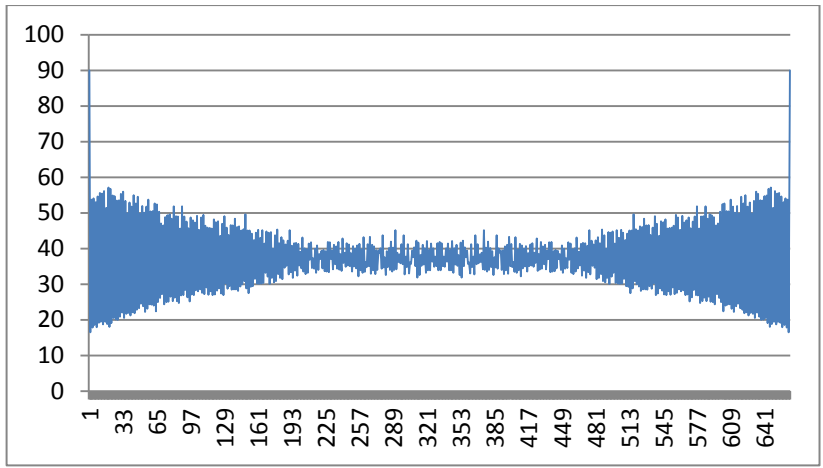

**Figure 3a.** Autocorrelation of Class A Baudhāyana sequences ordered by c

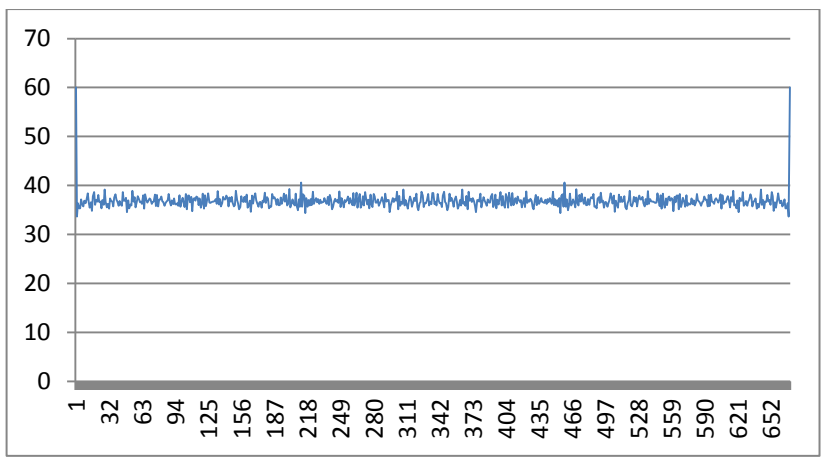

**Figure 3b**. Autocorrelation of Class B Baudhāyana sequences ordered by *c*

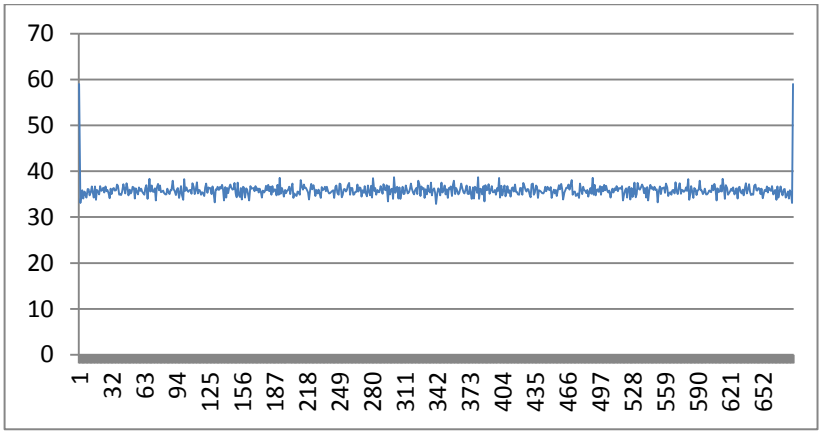

**Figure 3c**. Autocorrelation of Class C Baudhāyana sequences ordered by *c*

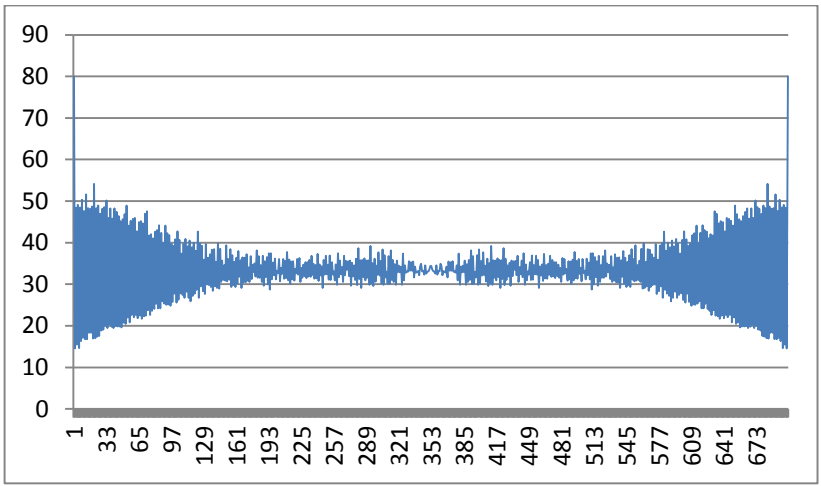

**Figure 3d**. Autocorrelation of Class D Baudhāyana sequences ordered by *c*

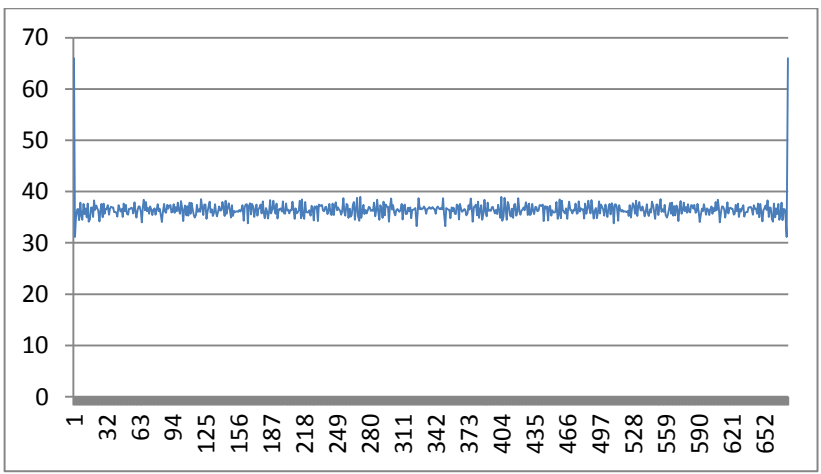

**Figure 3e**. Autocorrelation of Class E Baudhāyana sequences ordered by *c*

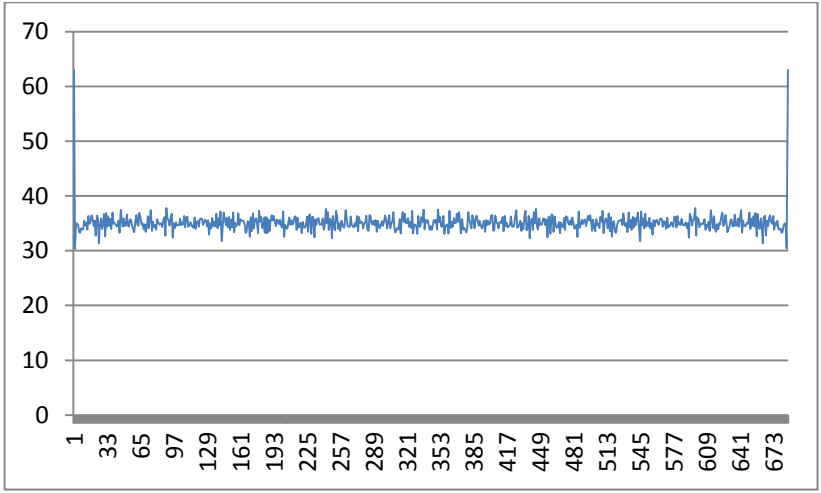

**Figure 3f.** Autocorrelation of Class F Baudhāyana sequences ordered by *c*

The A to F sequences may alternatively also be mapped into binary sequences and their properties remain qualitatively similar to the results of Figures 2 and 3. For example the autocorrelation of the binary sequence for Class E is shown in Figure 4 below.

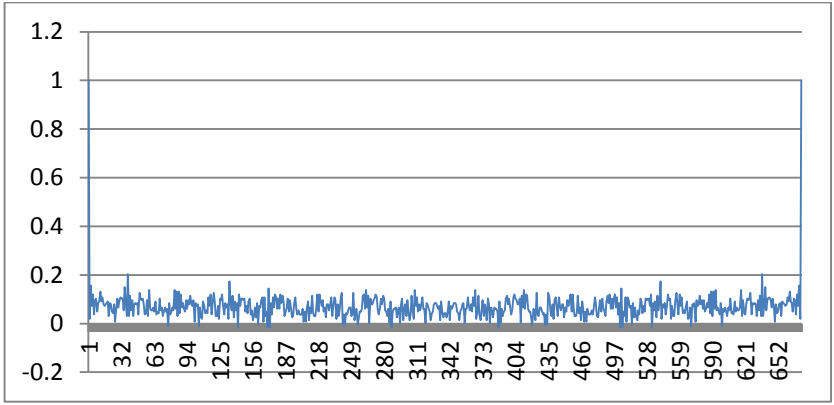

**Figure 4.** Autocorrelation of Class E for binary sequence

**Cross Correlation Properties of the Baudhāyana sequences**
Computation of the cross correlation functions reveals that the cross correlation is relatively high between A and D (Figure 5) when the Baudhāyana sequences are ordered according to *c* as is to be expected.

The cross correlation values between other pairs of Baudhāyana sequences are low (typically like that of Figure 6). This validates our assessment that the sequences possess excellent randomness properties.

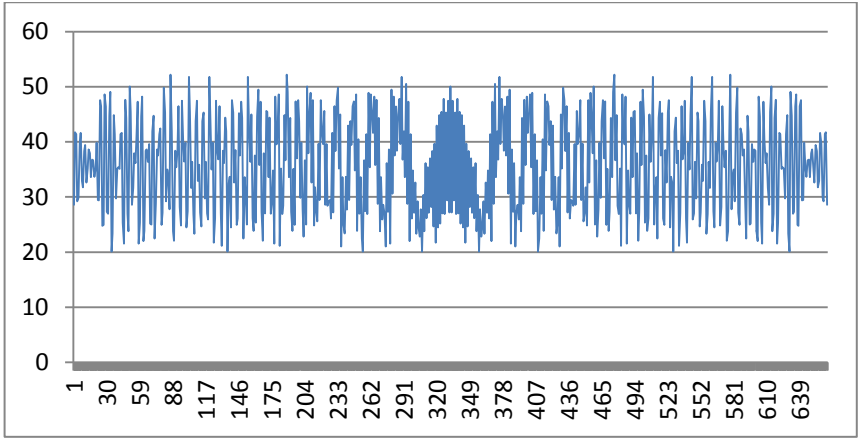

**Figure 5.** Cross correlation between A and D (the values range between 20 and 53)

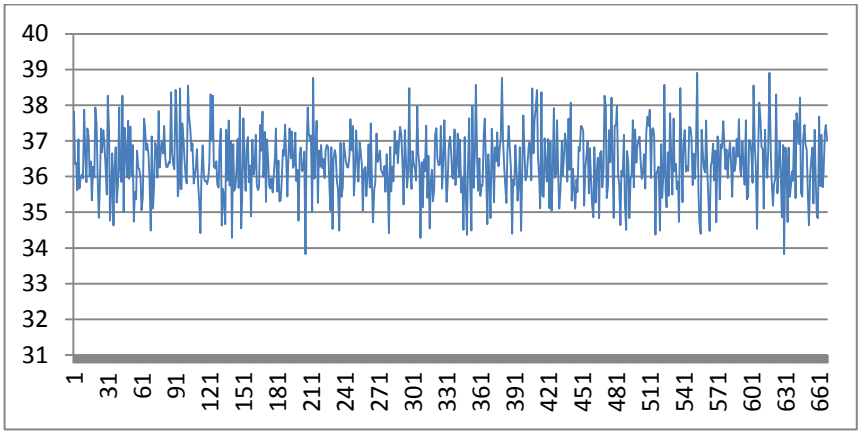

**Figure 6.** Cross correlation between B and C (the values range between 34 and 39)

**Conclusions**

This paper shows that the six classes of PPTs can be put into two larger classes. Specifically, autocorrelation and cross-correlation functions of the Baudhāyana sequences of the six classes have been computed. It is shown that Classes A and D (in which the largest term is divisible by 5) are different from the other four classes in their randomness properties if they are ordered by $c$. But if the Baudhāyana sequences are ordered by $a$ or $b$, each of the six classes exhibits excellent randomness properties. This remains true if binary mappings of the Baudhāyana sequences are considered.

We expect Baudhāyana sequences to have applications in key distribution and in information hiding [15]. Subsequences of these sequences may be indexed according to type (in terms of what number they have been ordered by) and class, the place in the larger sequence, and length, which are parameters that can be used in a cryptographic protocol.

**References**

1. S. Kak, Pythagorean triples and cryptographic coding. 2010. arXiv:1004.3770
2. S. Kak and A. Chatterjee, On decimal sequences. IEEE Transactions on Information Theory IT-27: 647-652, 1981.
3. S. Kak, Encryption and error-correction coding using D sequences. IEEE Transactions on Computers C-34: 803-809, 1985.
4. S.K.R. Gangasani, Analysis of prime reciprocal sequences in base 10. http://arxiv.org/abs/0801.3841v1
5. S. Kak, A two-layered mesh array for matrix multiplication. Parallel Computing 6: 383-385, 1988.
6. S. Kak, On the mesh array for matrix multiplication. 2010. arXiv:1010.5421
7. S. Rangineni, New results on scrambling using the mesh array. http://arxiv.org/abs/1102.4579v2
8. K.R. Kanchu and S. Kak, Goldbach circles and balloons and their cross correlation. arXiv:1209.4621
9. S. Kak, Random sequences based on the divisor pairs function. arXiv:1210.4614
10. F. J. M. Barning, On Pythagorean and quasi-Pythagorean triangles and a generation process with the help of unimodular matrices, (Dutch) Math. Centrum Amsterdam Afd. Zuivere Wisk, ZW-011, 1963.
11. D. McCullough, Height and excess of Pythagorean triples. Mathematics Magazine 78, 26-44, 2005.
12. H. Lee Price, The Pythagorean tree: a new species. arXiv:0809.4324v1
13. J.J. O'Conner and E.F. Robertson, Baudhayana. History of Mathematics Project. http://www-history.mcs.st-and.ac.uk/~history/Biographies/Baudhayana.html
14. A. Seidenberg, The origin of mathematics. Archive for History of Exact Sciences 18: 301-42, 1978.
15. M. Gnanaguruparan and S. Kak, Recursive hiding of secrets in visual cryptography. Cryptologia 26:68–76, 2002.